\documentclass{emulateapj}

\def\arcs{$''$}

\def\hub{\ifmmode H_\circ\else H$_\circ$\fi}

\shorttitle{Star clusters in M33} \shortauthors{Ma}

\begin{document}
\slugcomment{AJ, in press}
\title{An updated catalog of M33 clusters and candidates: $UBVRI$ photometry, and some statistical results}

\author{Jun Ma,\altaffilmark{1,2}}

\altaffiltext{1}{National Astronomical Observatories, Chinese Academy of Sciences, A20 Datun Road, Chaoyang District, Beijing 100012, China}

\altaffiltext{2}{Key Laboratory of Optical Astronomy, National Astronomical Observatories, Chinese Academy of Sciences, Beijing 100012, China}

\email{majun@nao.cas.cn}

\begin{abstract}
We present $UBVRI$ photometry for 392 star clusters and candidates in the field of M33, which are selected from the most recent star cluster catalog. In this catalog, the authors listed star clusters' parameters such as cluster positions, magnitudes and colors in the $UBVRIJHK_s$ filters, and so on. However, a large fraction of objects in this catalog do not have previously published photometry. Photometry is performed using archival images from the Local Group Galaxies Survey, which covers 0.8 deg$^2$ along the major axis of M33. Detailed comparisons show that, in general, our photometry is consistent with previous measurements. Positions (right ascension and declination)
for some clusters are corrected here. Combined with previous literature, we constitute a large sample of M33 star clusters. Based on this cluster sample, we present
some statistical results: none of the M33 youngest clusters ($\sim 10^7$~yr) have masses approaching $10^5$~$M_{\odot}$; roughly half the star clusters are
consistent with the $10^4$ to $10^5$~$M_{\odot}$
mass models; the continuous distribution of star clusters along the model line indicates that M33 star clusters have been formed continuously from the epoch of the first star cluster formation until recent times; there are $\sim 50$
star clusters which being overlapped with the Galactic globular clusters on the color-color diagram, and these clusters are old globular clusters candidates in M33.
\end{abstract}


\section{Introduction}
\label{s:intro}

The importance of the study of star clusters is difficult to overstate, especially in the Local Group galaxies. Star clusters, which represent, in distinct and luminous ``packets'', single-age and single-abundance points and encapsulate at least a partial history of the parent galaxy's evolution, can provide a unique laboratory for studying the ongoing and past star formation in the parent galaxy.

M33 is a small Scd Local Group galaxy, about 15 times
farther from us than the LMC (distance modulus $(m-M)_0=24.54\pm0.06$; McConnachie et al.
2004, 2005). It is interesting and important because it represents a morphological type intermediate between the
largest ``early-type'' spirals and the dwarf irregulars in the Local Group. So, it can provide an important link between the cluster populations of earlier-type spirals (Milky Way and M31) and the numerous, nearby later-type dwarf galaxies.

Although in the work of M31 globular clusters, \citet{hub32} remarked that he discovered some twenty or fifteen diffuse objects in M33 and remarked that they
averaged about 1.5 mag fainter than the globular clusters in M31, \citet{hiltner60} did the pioneering work of M33 star clusters. In his work, \citet{hiltner60}
used photographic plates taken with the Mt. Wilson 100 inch (2.5 m) telescope to photometer 23 M33 cluster candidates and 23 M31 globular clusters in the $UBV$ passbands, and found that, except for five ones, the clusters
in M33 are bluer and fainter than those in M31. At the same time, \citet{KM60} did photoelectric observations for 4 M33 star clusters. And then, \citet{MD78}
detected 58 star clusters in M33 based on a baked IIIa-J+GG385 plate which covering a field of about one degree in diameter. The most comprehensive
catalog of nonstellar objects in M33 was compiled by \citet{CS82,CS88}, who detected 250 nonstellar objects by visually examining a single photographic plate taken at the Ritchey-Chrestien focus of the 4 m telescope at Kitt Peak National Observatory, and obtained ground-based $B$, $V$, and $I$ photometry of 106 of these objects, for which they believe to be star clusters. Recently, \citet{Mochejska98}
detected 51 globular cluster candidates in M33, 32 of
which were not previously cataloged, using the data
collected in the DIRECT project \citep{Kaluzny98,Stanek98}. Since the pioneering work of \citet{CBF99a}, the era of detecting and studying M33 star clusters based on the images with {\sl Hubble Space Telescope} ({\sl HST}) has begun \citep{CBF99a,CBF99b,CBF99c,CBF01,CBF02,Bedin05,PL07,Sarajedini07,
Stonkute08,Huxor09,Roman09,ZK09}.
\citet{Ma01,Ma02a,Ma02b,Ma02c,Ma04a,Ma04b} constructed spectral energy distributions (SEDs) in 13 intermediate filters of the Beijing-Arizona-Taiwan-Connecticut (BATC) photometric system for known M33 clusters and candidates from \citet{MD78}, \citet{CBF99a,CBF01} and \citet{Mochejska98}, and estimated cluster properties. In order to construct a single master catalog incorporating the entries in all of the individual catalogs including all known properties of each cluster, \citet{sara07} merged all of the above-mentioned catalogs before 2007, for a summary of the properties of all of these catalogs. This catalog
contains 451 candidates, of which 255 are confirmed clusters based on {\sl HST} and high-resolution ground-based imaging. The positions of the clusters in \citet{sara07} were transformed to the J2000.0 epoch and refined using the Local Group Galaxies Survey
(LGGS; Massey et al. 2006). In addition, some authors used the images observed with the MegaCam camera on the 3.6 m Canada-France-Hawaii Telescope (CFHT/MegaCam) to detect star clusters in M33 \citep{ZK08,Roman10};
\citet{Sharina10} presented the evolutionary parameters of 15 GCs in M33 based on the results of medium-resolution spectroscopy obtained at the Special Astrophysical Observatory 6-m telescope. Most recently, \citet{Robert11} search for outer halo star clusters in M33 based on CFHT/MegaCam imaging as part of the Pan-Andromeda Archaeological Survey (PAndAS).

As \citet{sara07} pointed out that, the photometry of M33 clusters and cluster candidates are from the various original catalogs which are all on different zero points. In addition, more than 160 clusters and cluster candidates do not possess any photometric data in the M33 adopted cluster catalog of \citet{sara07}. So, it is important to provide photometry for these clusters and cluster candidates of M33 which being without any photometric data, and it is also important to provide photometry for clusters and cluster candidates of M33 in the same photometric system.

In this paper, we perform aperture photometry of 392 M33 star clusters and cluster candidates based on the LGGS images of M33. These sample clusters are selected from the M33 adopted cluster catalog of \citet{sara07}. This paper is organized as follows. \S 2 describes the sample selection and $UBVRI$ photometry. In \S 3, we present an analysis of the cluster properties. Lastly, our conclusions are presented in \S 4.

\section{Data}
\label{s:data}

\subsection{Sample}
\label{s:samp}

We selected our sample star clusters and cluster candidates from the M33 adopted cluster catalog of \citet{sara07}, which is a compilation of photometry and identifications from many previous catalogs. This catalog contains precise cluster positions (right ascension and declination), magnitudes and colors in the $UBVRIJHK_s$ filters, metallicities, radial velocities, masses and ages, where available, and galactocentric distances for each cluster. However, from this catalog, we can see that, for more than 160 objects there are not any photometric data. So, homogeneous photometric data are urgently needed. We used archival $UBVRI$ images of M33 from the LGGS available from their ftp site \footnote{ftp://ftp.lowell.edu/pub/massey/lgsurvey/datarelease/}, which covers a region of 0.8 deg$^2$ along the galaxy's major axis. The images we used consist of 3 separate but overlapping fields with a scale from 0.261\arcs pixel$^{-1}$ at the center to 0.258\arcs pixel$^{-1}$ in
the corners of each image. The field of view of each mosaic image is $36'\times36'$. The observations were taken from August 2000 to September 2002 with the KPNO 4 m telescope. The median seeing of the LGGS images is $\sim$ 1\arcs. We employed {\sc iraf/daofind} to find the sources in
the images and match them to the coordinates of the M33 adopted cluster catalog of \citet{sara07}. In this paper, we will perform photometry for star clusters and cluster candidates in the M33 adopted cluster catalog of \citet{sara07}, in which there are 393 star clusters and cluster candidates. To prevent mistakes, we checked each object visually in the images. Except for 3 objects (8, 287 and 417), the coordinates presented by \citet{sara07} are of sufficient accuracy to make the objects be easily discernible. For objects 8 and 287, the offsets of the coordinates presented by \citet{sara07} and this paper are $\sim 3.0''$ and $\sim 1.3''$ (For object 8, its R.A. and REC are 01:32:41.27 and +30:27:51.9~(J2000.0) presented by Sarajedini \& Mancone [2007] compared to 01:32:41.273 and +30:27:54.76~(J2000.0) given here; for object 287, its R.A. and REC are 01:34:03.34 and +30:48:28.0 presented by Sarajedini \& Mancone [2007] compared to 01:34:03.311 and +30:48:26.73~(J2000.0) given here), respectively. For object 417, i.e. U139 of \citet{CS82}, its R.A. and REC listed by \citet{sara07} are 01:34:36.92 and +30:03:47.6~(J2000.0), which falls the outside of the region covered by the LGGS images, however, its R.A. and REC listed by \citet{CS82} are 01:29:47 and +30:12:44~(J1950.0), i.e. 01:32:35.88 and +30:28:07.96~(J2000.0). The offset of these two sets of coordinates given by \citet{CS82} and \citet{sara07} is too large. So, the R.A. and REC of object 417 presented by \citet{sara07} may be typing error. Based on the original R.A. and REC listed by \citet{CS82}, we found that only one object exits within $20''$, so it is reasonable that this object is U139 of \citet{CS82},
whose R.A. and REC derived based on the LGGS images are 01:32:35.786 and +30:28:09.16~(J2000.0). In addition, according to the R.A. and REC presented by \citet{sara07}, object 268 and 269 is the same source. We delete number 268. So, the last sample of this paper includes 392 star clusters and cluster candidates of M33, which are selected from the M33 adopted cluster catalog of \citet{sara07}. Figure 1 shows the spatial distribution of the 392 objects in the LGGS fields. The large ellipse is the $D_{25}$ boundary of the M33 disk \citep{Vaucouleurs91}. The 3 large squares are the LGGS field boundaries.

\begin{figure*}
\centerline{\includegraphics[scale=0.8,angle=-90]{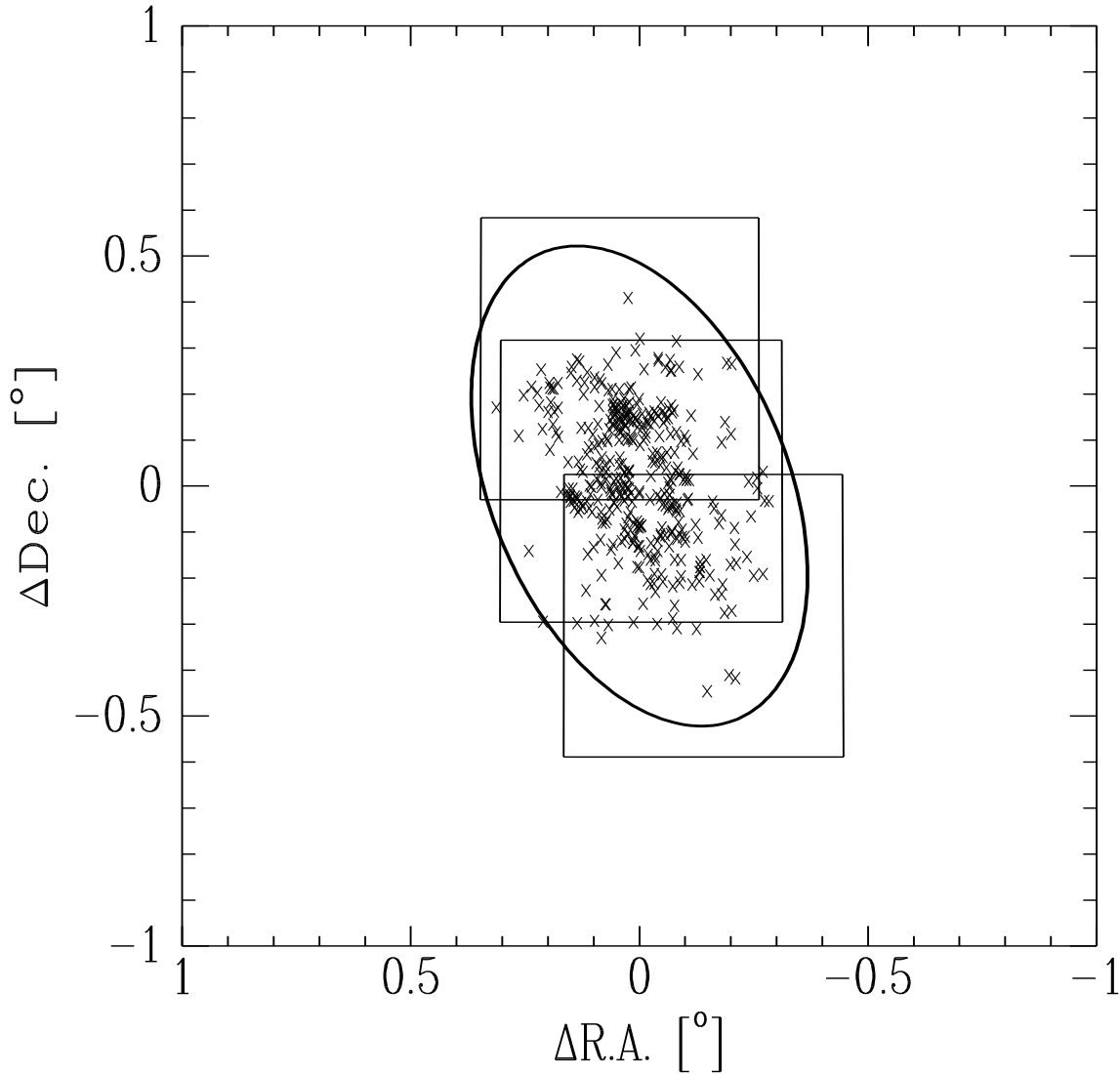}}
\caption[]{Spatial distribution of the 392 star clusters and cluster candidates of M33 which being selected from \citet{sara07} and their loci in the LGGS fields.
We determined the photometry for these objects based on the LGGS archival images of M33 in the $UBVRI$ bands.
The large ellipse is the $D_{25}$ boundary of the M33 disk \citep{Vaucouleurs91}. The three large squares are the LGGS fields.}
\end{figure*}

\subsection{Integrated photometry}
\label{s:phot}

We used the LGGS archival images of M33 in the $UBVRI$ bands to perform photometry. Previously, \citet{massey} compiled point-spread-function (PSF) photometry for 146,622 stars (point sources) in the M33 fields, with photometric uncertainties of $<10$\% below $V = 23$ mag. However, there is as yet no published LGGS photometry for extended sources, such as star clusters and galaxies.

We performed aperture photometry of these 392 M33 star clusters and cluster candidates found in the LGGS images in all of the $UBVRI$ bands to provide a comprehensive and homogeneous photometric catalog for these objects. The photometry routine we used is {\sc iraf/daophot} \citep{stet}. To determine the total luminosity of each cluster, we produced curve of growth from $V$-band photometry obtained through apertures with radii in the range 3-40 pixel with 1 pixel increments. These were used to determine the aperture size required to enclose the total cluster light. The most appropriate photometric radius that includes all light from the objects, but excludes (as much as possible and to the extent that this was obvious) extraneous field stars is adopted. Figure 2 shows curves of growth for 8 clusters selected randomly according to luminosity. In Figure 2, the most appropriate photometric radius needed for photometry is indicated by  triangles. In addition, we have checked the aperture of every sample object considered here by visual examination to make sure that it was large enough to include all light from this object, but not too large (to avoid contamination from other sources). The local sky background was measured in an annulus with an inner radius which being larger 1 pixel than photometric radius and 5 pixels wide, in which the mode was used. The instrumental magnitudes were then calibrated to the standard Johnson-Kron-Cousins $UBVRI$ system by comparing the published magnitudes of stars from \citet{massey}, who calibrated their photometry with standard stars of \citet{lan92}, with our instrumental magnitudes. Since the magnitudes in \citet{massey} are given in the Vega system, our photometry is also tied to that system. Finally, except for object 216 in $U$ band, which falls in the gap of the image, we obtained photometry for 392 objects in the individual $UBVRI$ bands. Table 1 lists our new $UBVRI$ magnitudes and the aperture radii used (We adopted 0.258\arcs pixel$^{-1}$ from the image header.), with errors given by {\sc iraf/daophot}. The object names follow the naming convention of \citet{sara07}.

\begin{figure*}
\centerline{
\includegraphics[height=140mm,angle=-90]{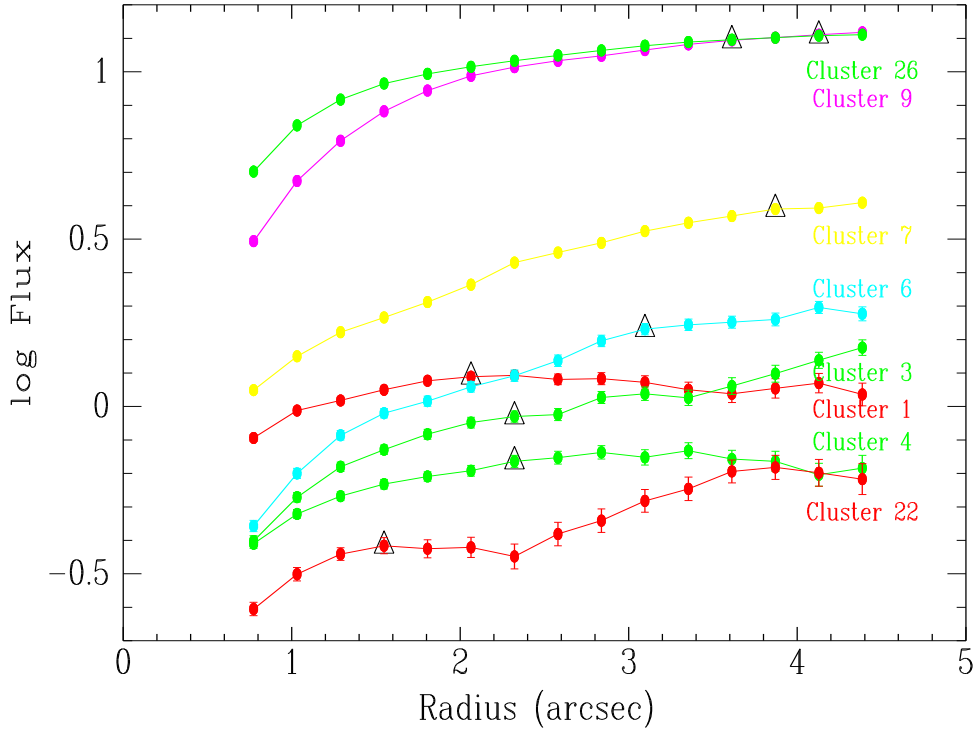}}
\vspace{0.5cm}
\caption{Curves of growth for 8 clusters of M33 selected randomly according to luminosity. Triangles indicate the radii of the apertures needed for the photometry.}
\end{figure*}

To examine the quality and reliability of our photometry, we compared the aperture magnitudes of the 392 objects considered here with the magnitudes collected from various sources in \citet{sara07}, and with previous measurements in \citet{PL07} and \citet{Roman09}. There are 18 clusters, of which the magnitude scatters in $V$ band between this and previous studies \citep{sara07,PL07,Roman09} are larger than 1.0 mag, i.e. our aperture magnitudes are fainter than previous measurements in \citet{sara07}, \citet{PL07} and \citet{Roman09}. We listed these objects in Table 2, and we also plotted images of these objects in Figure 3. The circles are photometric apertures adopted here. From this figure, we can see that most of these objects (44, 45, 66, 116, 118, 153, 195, 221, 231, 250, 253, 276, 338, and 367) are very close to one or more bright sources. If photometric apertures are larger than the values adopted here, the light from these bright sources will not be excluded. The object 46 is really faint. As we know, for objects in crowed fields or for faint objects, different aperture sizes adopted for photometry would cause a large scatter in the photometric measurement. In fact, from Table 2, we can see that, for some star clusters (45, 66, 193, 195, 221, 231, 250, and 367), the photometric measurements obtained by \citet{PL07} are also very different from the photometric data collected by \citet{sara07}. In \citet{sara07}, the photometric data, which are collected from various original catalogs and are on different zeropoints, are transformed to the reference system of \citet{CBF99a,CBF01} by applying offsets derived from objects in common between the relevant catalog and the data set of \citet{CBF99a,CBF01}. \citet{CBF99a,CBF01} derived $UBV$ or $BVI$ photometry of M33 star clusters based on images taken with {\sl HST}/WPC2 with an aperture of $r=2.2''$ for $V$ magnitude measurement and an aperture of $r=1.0''$ for the measurement of color. In \citet{PL07}, $BVI$ integrated aperture photometry of M33 star clusters, which are included in $50'\times80'$ field of M33 based on CCD images taken with the CFH12k mosaic camera at the CFHT, are derived with an aperture of $r=4.0''$ for $V$ magnitude measurement and an aperture of $r=2.0''$ for the measurement of color. \citet{Roman09} derived integrated photometry and color-magnitude diagrams (CMDs) for 161 star clusters in M33, using the Advanced Camera For Surveys (ACS) Wide Field Channel (WFC) onboard the {\sl HST}. These authors adopted an aperture radius of $r=2.2''$ for $V$ magnitude measurements and $r=1.5''$ for the colors. For these 18 sources, the large magnitude scatters in $V$ band between this and previous studies \citep{sara07,PL07,Roman09} come from different photometric aperture sizes adopted by different authors.

Figures 4, 5 and 6 show the comparison of our photometry of the clusters considered here with previous photometric data in \citet{sara07}, with previous photometric measurements in \citet{PL07} and \citet{Roman09}. Objects 46, 116, and 193 are not included in the figure of $\Delta V$ comparison of Figure 4 because of too large values of $\Delta V$ to be drawn in the figure. The photometric offsets and rms scatter of the differences between previous measurements and our new magnitudes are summarized in Tables 3, 4 and 5.

\begin{figure*}
\centerline{
\includegraphics[scale=0.8,angle=0]{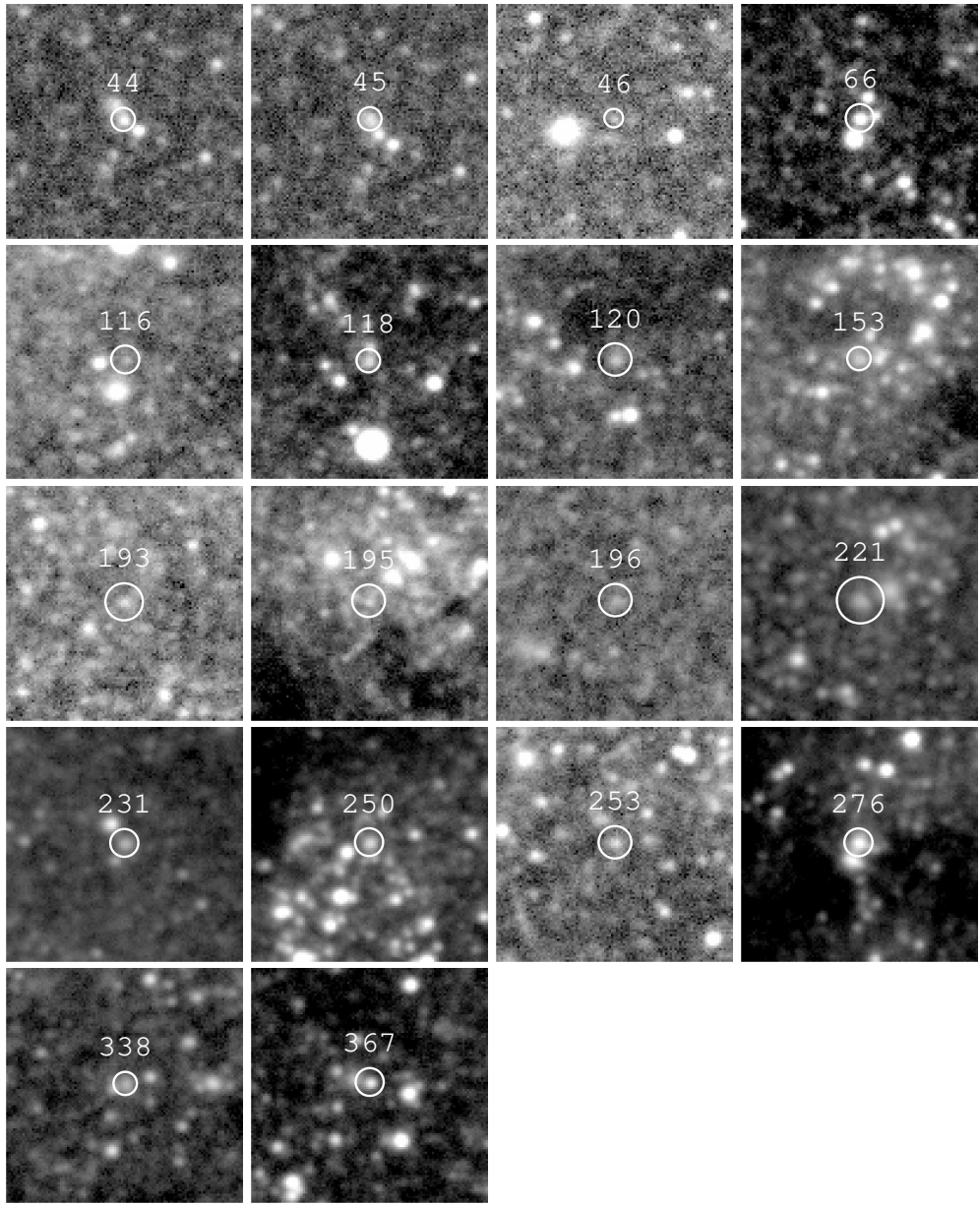}}
\vspace{-4cm}
\caption{Finding charts of 18 star clusters and candidates of M33 in the LGGS $V$ band, of which the magnitude scatters in $V$ band between this and those studies of \citet{sara07}, \citet{PL07} and \citet{Roman09} are larger than 1.0 mag, i.e. our measurements are fainter than those
in \citet{sara07}, \citet{PL07} and \citet{Roman09}. The circles are photometric apertures adopted in this paper.}
\end{figure*}

\begin{figure*}
\centerline{
\includegraphics[height=140mm,angle=-90]{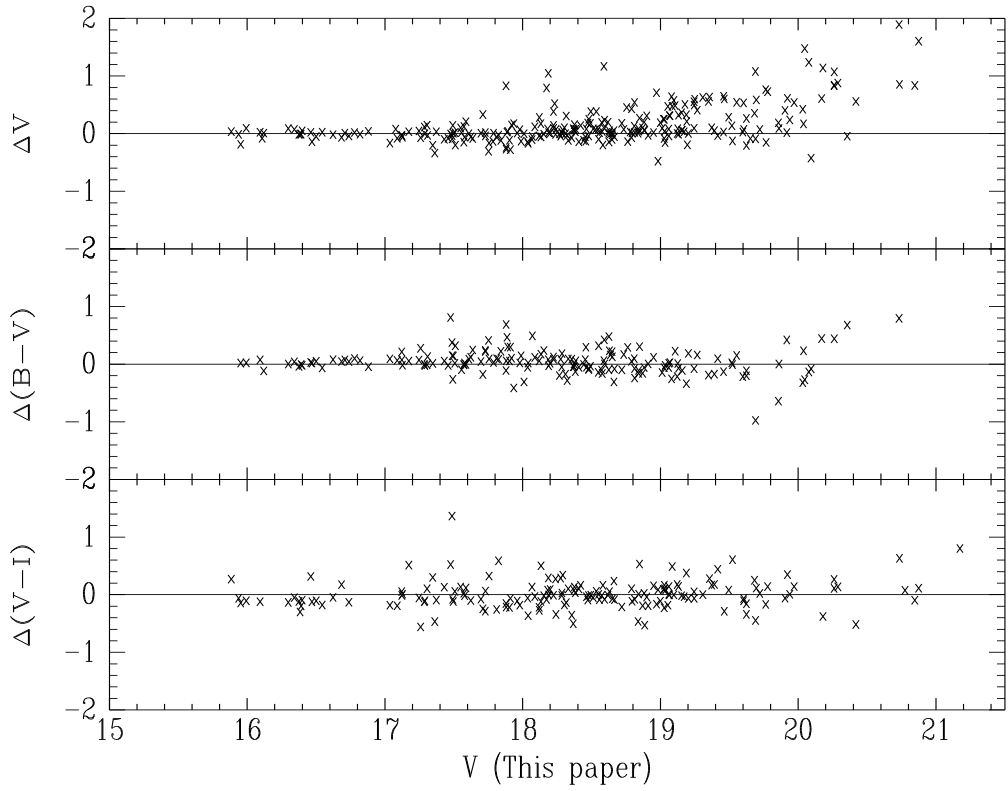}}
\vspace{0.5cm}
\caption{Comparisons of our photometry of M33 star clusters and candidates in the $UBVRI$ bands with previous measurements being collected in \citet{sara07}.}
\end{figure*}

\begin{figure*}
\centerline{
\includegraphics[height=140mm,angle=-90]{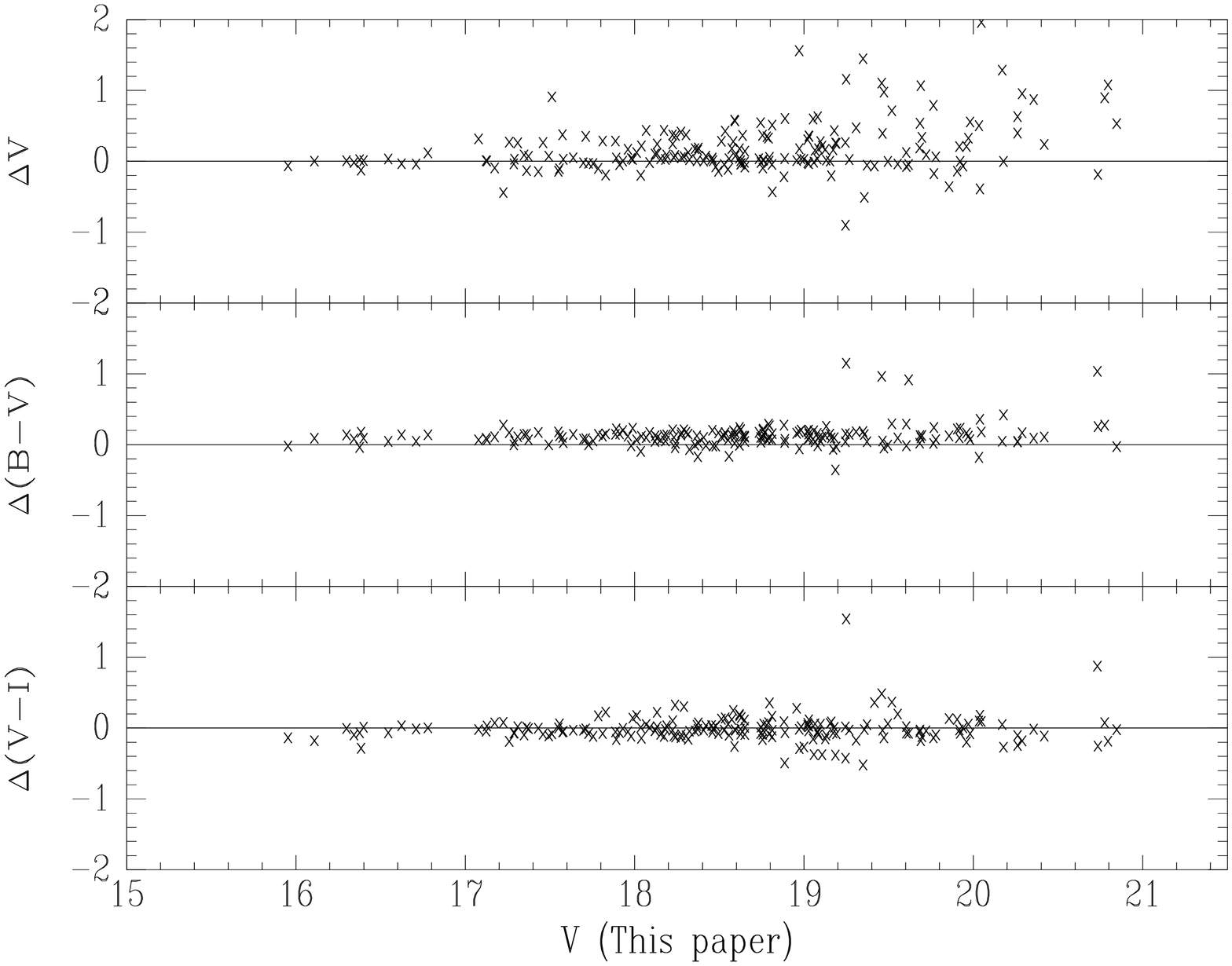}}
\vspace{0.5cm}
\caption{Comparisons of our photometry of M33 star clusters and candidates in the $UBVRI$ bands with previous photometry in \citet{PL07}.}
\end{figure*}

\begin{figure*}
\centerline{
\includegraphics[height=140mm,angle=-90]{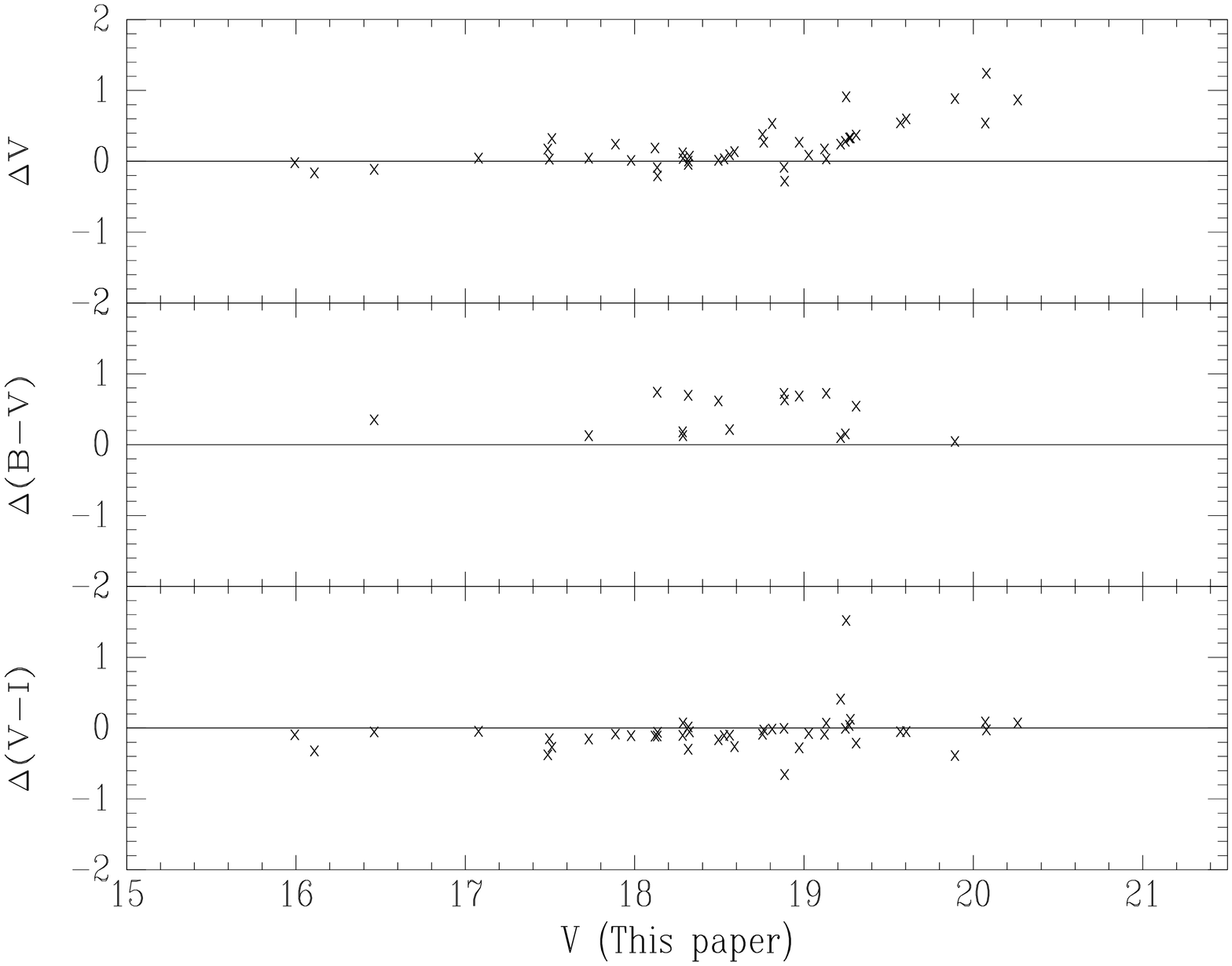}}
\vspace{0.5cm}
\caption{Comparisons of our photometry of M33 star clusters and candidates in the $UBVRI$ bands with previous photometry in \citet{Roman09}.}
\end{figure*}

From Figures 4, 5 and 6 and Tables 3, 4 and 5, we can see that, our measurements in $V$ band get systematically fainter than the photometric data in \citet{sara07}, and the photometric measurements in \citet{PL07} and \citet{Roman09} for fainter sources ($V\geq19$). Except for the $(B-V)$ difference between this study and those of \citet{PL07} and \citet{Roman09}, which turned out to be $0.127\pm 0.013$ with $\sigma=0.161$ and $0.428\pm 0.066$ with $\sigma=0.256$, both the $(B-V)$ and $(V-I)$ colors obtained here are in good agreement with those in \citet{sara07}, \citet{PL07} and \citet{Roman09}. For the $(B-V)$ colors obtained by \citet{Roman09}, we will discuss them in \S 3.2 and 3.3.

To check whether and how seriously aperture variations affect the photometric measurements, we performed test with the same size aperture for all objects considered here. We chose $3.354''$ (corresponding to 13 pixels in the LGGS images) as a radius for obtaining photometric magnitudes of objects considered here. This aperture size is nearly in between $2.2''$ and $4.0''$, which are chosen to obtain $V$ magnitude measurements of M33 star clusters by \citet{CBF99a,CBF01} and \citet{Roman09}, and \citet{PL07}, respectively. Figures 7, 8 and 9 show the comparison of our photometry of the clusters considered here with previous photometric data collected by \citet{sara07}, and with previous photometry of \citet{PL07} and \citet{Roman09}. Tables 6, 7 and 8  summarize photometric offsets and rms scatter of the differences between previous measurements and our new magnitudes. It is evident that the photometries obtained with a radius of $3.354''$ agree good with those in \citet{sara07}, \citet{PL07} and \citet{Roman09}. From Figures 7, 8 and 9 and Tables 6, 7 and 8, we can see that, except for the $(B-V)$ difference between our study and those of \citet{PL07} and \citet{Roman09}, which turned out to be $0.107\pm 0.011$ with $\sigma=0.154$ and $0.405\pm 0.068$ with $\sigma=0.263$, both the $(B-V)$ and $(V-I)$ colors and $V$ magnitudes obtained here are in good agreement with previous photometric measurements in \citet{sara07}, \citet{PL07} and \citet{Roman09}, and there are no evident difference between the photometric zeropoint here as compared with \citet{sara07}, \citet{PL07} and \citet{Roman09}. There are 10 clusters, of which the magnitude scatters in $V$ band between this
study and those of \citet{sara07}, \citet{PL07} and \citet{Roman09} are larger than one magnitude, i.e. our photometric measurements are fainter or brighter than those in \citet{sara07}, \citet{PL07} and \citet{Roman09}. By comparing Table 2 with Table 9, we find that, except for object 196, the $V$ magnitude of which is nearly not dependent on aperture size, the $V$ magnitudes of other common objects (46, 193, 195, and 231 ) decrease with aperture sizes. For objects 90 and 132 in Table 9, the $V$ magnitudes obtained with an aperture radius of $r=3.354''$ here are all brighter than previous measurements, however, for object 426 in Table 9, the photometric magnitude obtained with an aperture radius of $r=3.354''$ here is still fainter than previous measurements in \citet{sara07} and \citet{PL07}. For objects 132, 193, 231, 240 and 426, our photometric measurements are in good agreement with those of \citet{PL07}. In addition, for object 236, our photometric measurement is in agreement with that of \citet{sara07}, but is brighter than that of \citet{PL07}. We also note that, in Table 9, there are only 3 objects (90, 195, and 236), of which the $V$ magnitude scatters between this study and that of \citet{PL07} are larger than 1.0 mag.

\begin{figure*}
\centerline{
\includegraphics[height=140mm,angle=-90]{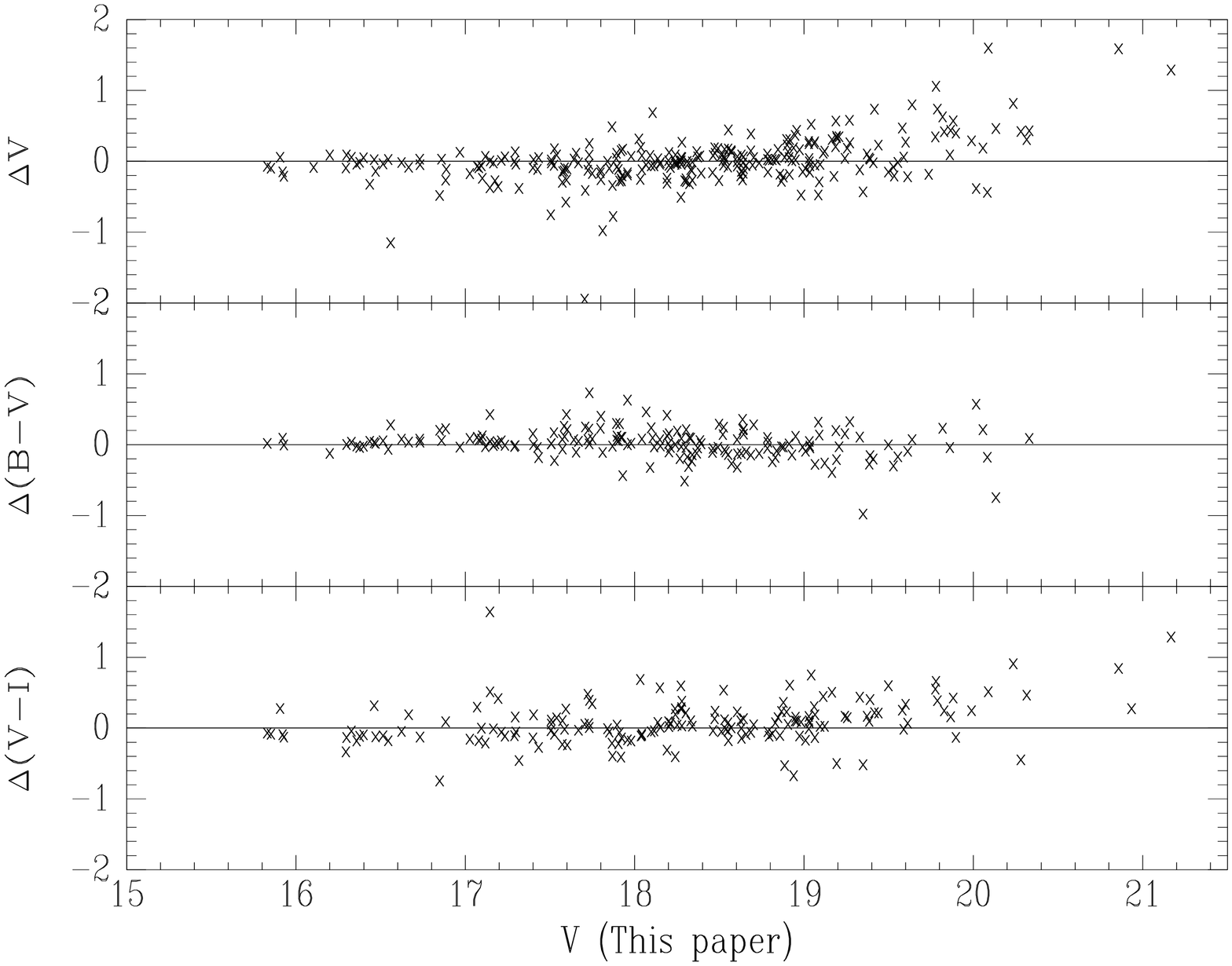}}
\vspace{0.5cm}
\caption{Comparisons of our photometry of M33 star clusters and candidates in the $UBVRI$ bands with previous measurements being collected in \citet{sara07}. Photometries of M33 star clusters are derived with an aperture of $r=3.354''$ (13 pixels) in this paper.}
\end{figure*}

\begin{figure*}
\centerline{
\includegraphics[height=140mm,angle=-90]{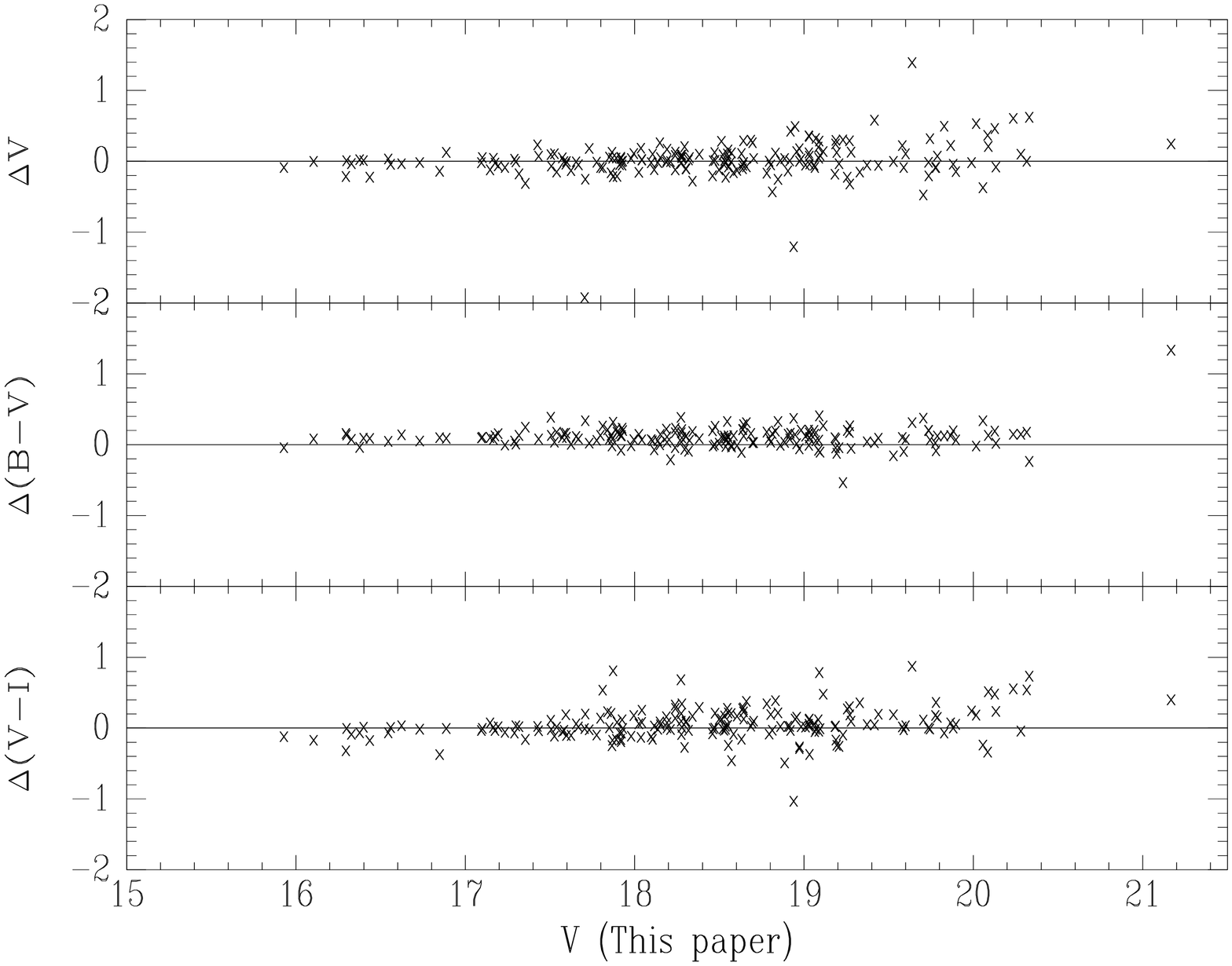}}
\vspace{0.5cm}
\caption{Comparisons of our photometry of M33 star clusters and candidates in the $UBVRI$ bands with previous photometry in \citet{PL07}. Photometries of M33 star clusters are derived with an aperture of $r=3.354''$ (13 pixels) in this paper.}
\end{figure*}

\begin{figure*}
\centerline{
\includegraphics[height=140mm,angle=-90]{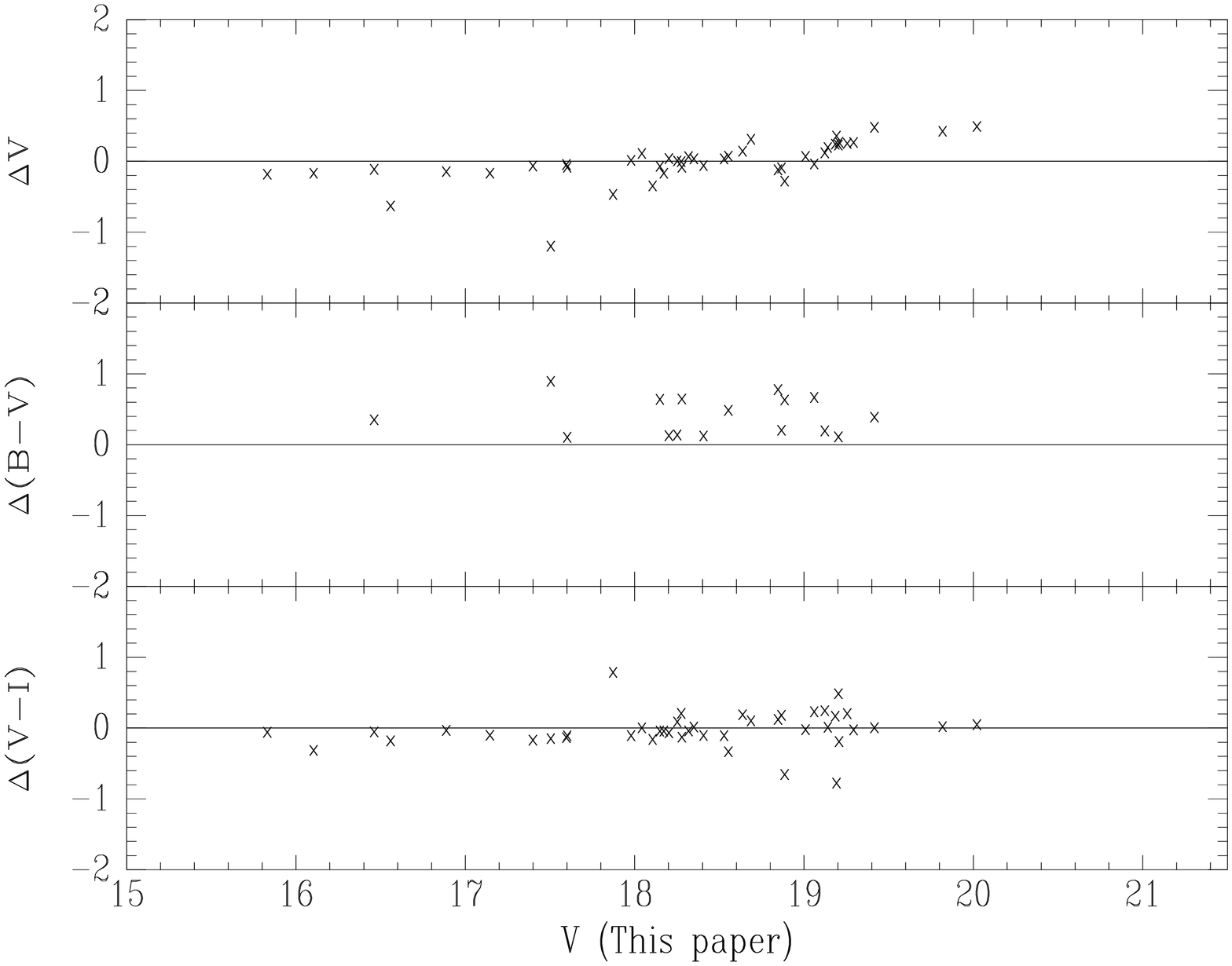}}
\vspace{0.5cm}
\caption{Comparisons of our photometry of M33 star clusters and candidates in the $UBVRI$ bands with previous photometry in \citet{Roman09}. Photometries of M33 star clusters are derived with an aperture of $r=3.354''$ (13 pixels) in this paper.}
\end{figure*}

\section{Statistical properties of M33 star clusters}

\subsection{Sample}

Our statistical sample contains 521 confirmed M33 star clusters, of which 254 have $UBVRI$ photometry obtained here, 35 have $BVI$ photometry obtained by \citet{PL07}, 117 have $BVI$ photometry obtained by \citet{Roman09}, and 115 have $BVI$ photometry obtained by \citet{ZK09}. These star clusters were confirmed based on the {\sl HST} or high-resolution ground-based imaging \citep[see][for details]{sara07,PL07,Roman09,ZK09}. Figure 10 shows the spatial distribution of the 521 confirmed M33 star clusters. The large ellipse is the $D_{25}$ boundary of the M33 disk \citep{Vaucouleurs91}.

\begin{figure*}
\centerline{
\includegraphics[scale=0.8,angle=-90]{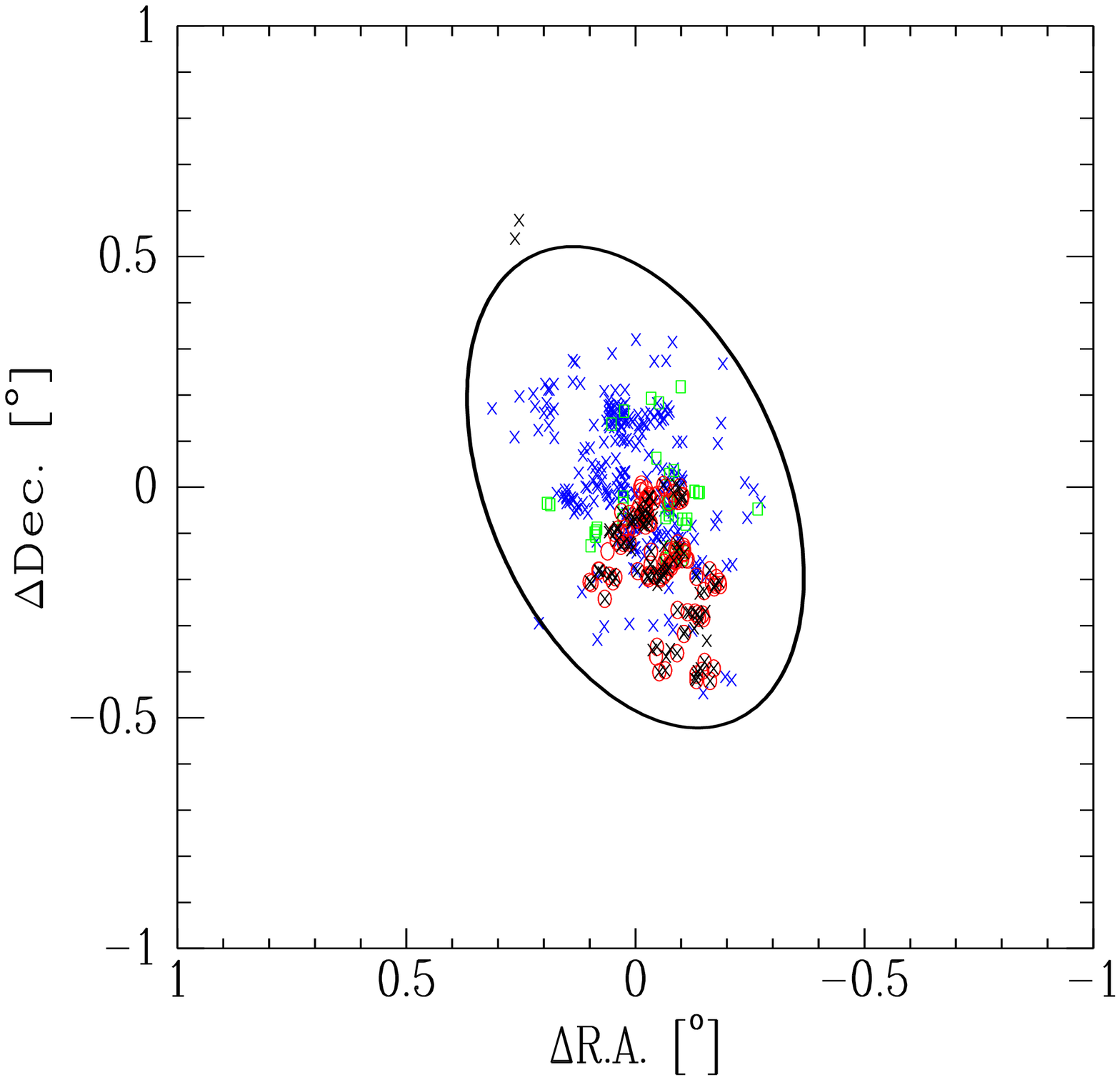}}
\caption[]{Spatial distribution of the 406 confirmed stat clusters in M33. Blue crosses denote the star clusters in this study, green squares denote the star clusters in
\citet{PL07}, red open circles denote the star clusters in \citet{Roman09}, and black crosses denote the star clusters in \citet{ZK09}. The large ellipse is the $D_{25}$ boundary of the M33 disk \citep{Vaucouleurs91}.}
\label{fig10}
\end{figure*}

\subsection{Color-magnitude diagram}

The color-magnitude diagram (CMD) provides a qualitative
model-independent global indication of cluster-formation
history that can be compared between galaxies, because $(B-V)_0$ and $(V-I)_0$ are reasonably good age indicators, at least between young and old populations, with a secondary dependence on metallicity \citep{CBF99b}. CMDs of M33 clusters have been previously discussed in
the literature \citep{CS82,CS88,CBF99b,PL07}. However, with a much larger cluster sample, it is worth investigating them again. Figure 11 displays the integrated $M_V-(B-V)_0$ and $M_V-(V-I)_0$ CMDs of the sample star clusters of M33.
The absolute magnitudes of the star clusters were derived for the adopted distance modulus of $(m-M)_0 = 24.69$ obtained by \citet{Galleti04}. The interstellar extinction curve, $A_{\lambda}$, is taken from \citet{car89}, $R_{V}=A_V/E(B-V)=3.1$. For reddening values of the star clusters, we used those in \citet{PL07} and \citet{Roman09}. For star clusters, \citet{PL07} and \citet{Roman09} did not present their reddening values, we adopted a uniform value of $E(B-V)=0.1$, as typical of the published values for the line-of-sight reddenings to M33 as \citet{sara07} adopted. Below each CMD in Figure 11 we have plotted the cluster distribution in color space. To the right of each CMD in Figure 11 we have shown a histogram of the clusters' absolute $V$ magnitudes.

From Figure 11, we can see that, the star clusters are roughly separated into blue and red groups with a color boundary of $(B-V)_0\simeq0.5$ in the $M_V-(B-V)_0$
CMD (vertical dashed line). This feature is also found by \citet{PL07} with a smaller sample. In fact,
Figure 6 of \citet{sara07} also shows this color separation. However, this separation in the $M_V-(V-I)_0$ CMD is not as clear as in the $M_V-(B-V)_0$
CMD, which was also shown by \citet{PL07}. Figure 11
shows that the cluster luminosity function peaks near $M_V\sim -6.0$ mag. In addition, there are some very red star clusters, the $(B-V)_0$ colors of which are much redder than 1.0 mag: $(B-V)_0=1.959$, 1.679, and 1.495 for clusters 116, 279, and 367 according to the photometry here. The $(B-V)_0$ colors of clusters 279 and 367 which being named 89 and 214 in \citet{PL07}, also derived by \citet{PL07}: 1.597 and 0.521. There is one star cluster, the $(B-V)_0$ color of which is much bluer than $-0.5$ mag: $(B-V)_0=-0.840$ for 111 of \citet{Roman09} according to the photometry of \citet{Roman09}. We did not derive the photometry for this star cluster. For the $(V-I)_0$ colors, there is one star cluster with $(V-I)_0=3.109$: cluster 216 of \citet{sara07} according to the photometry here. There is not previous photometry for cluster 216. From Figure 11, we also see that, the faintest star clusters identified so far in M33 are from \citet{ZK09}, there are about eight star clusters fainter than $-4.0$ mag in $V$ band.

By adding models to the CMDs, we can obtain a more detailed history of cluster formation. Three fading lines ($M_V$ as a function of age) of \citet{bc03} for a metallicity of $Z=0.004, Y=0.24$ which being thought to be appropriate for M33 star clusters \citep{CBF99b}, assuming a Salpeter initial mass function \citep{salp55} with lower and upper-mass cut-offs of $m_{\rm L}=0.1~M_{\odot}$ and $m_{\rm U}=100~M_{\odot}$, and using the Padova-1994 evolutionary tracks, are plotted on the CMDs of
M33 star clusters for three different total initial masses: $10^5$, $10^4$, and $10^3$~$M_{\odot}$. The majority of M33 star clusters fall between these three fading lines. From Figure 11, we note that none of the youngest clusters ($\sim 10^7$~yr) have masses approaching $10^5$~$M_{\odot}$, which is consistent with the results of \citet{CBF99b}. The fading lines show that, qualitatively, roughly half the star clusters here are consistent with the $10^4$ to $10^5$~$M_{\odot}$ mass models. For ages older than $10^9$ yr, some clusters with substantially higher masses are seen. From Figure 11, we can see that, some $(B-V)$ colors obtained by \citet{Roman09} are not consistent with the SSP lines.

\begin{figure*}
\centering
\includegraphics[height=135mm,angle=-90]{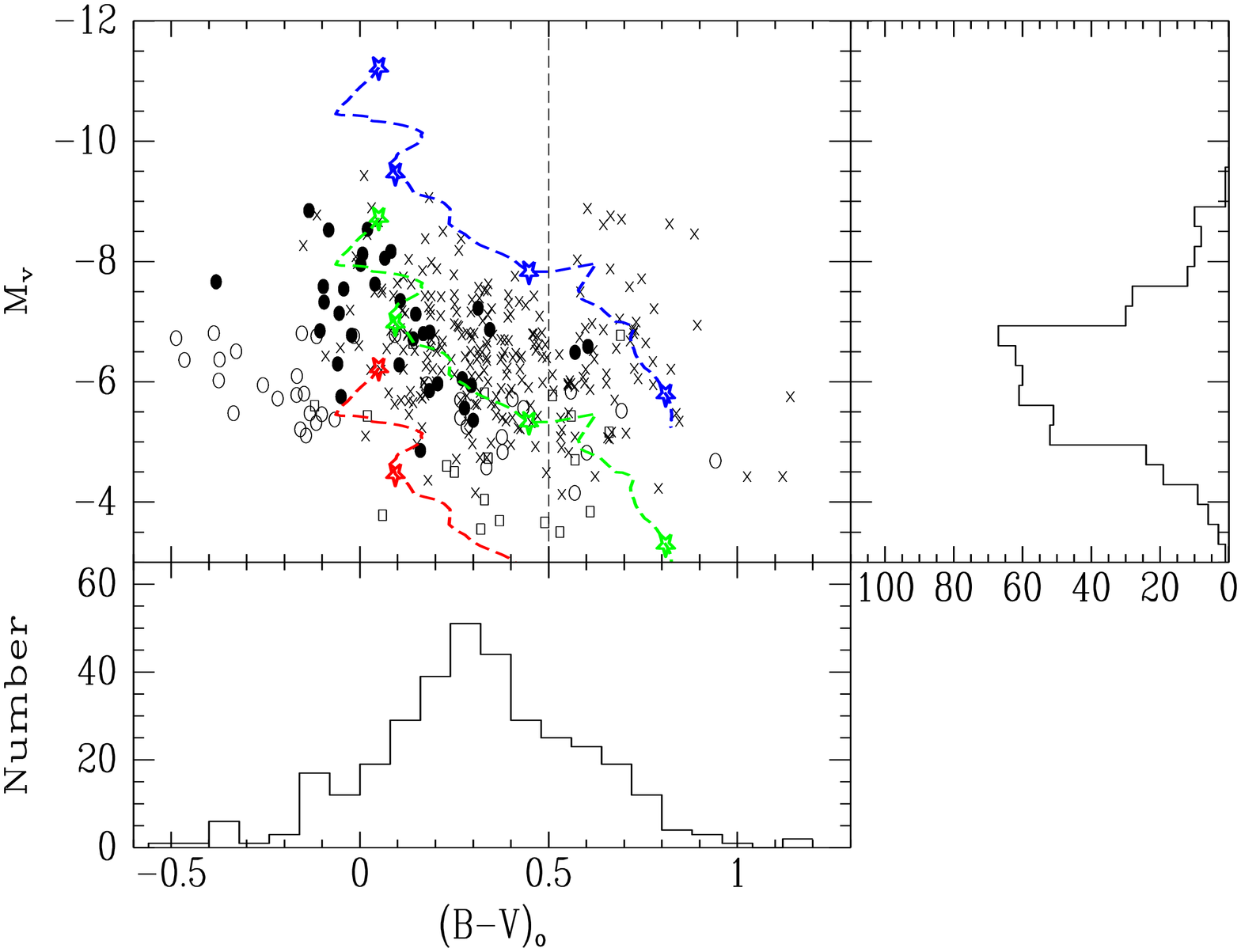}
\includegraphics[height=135mm,angle=-90]{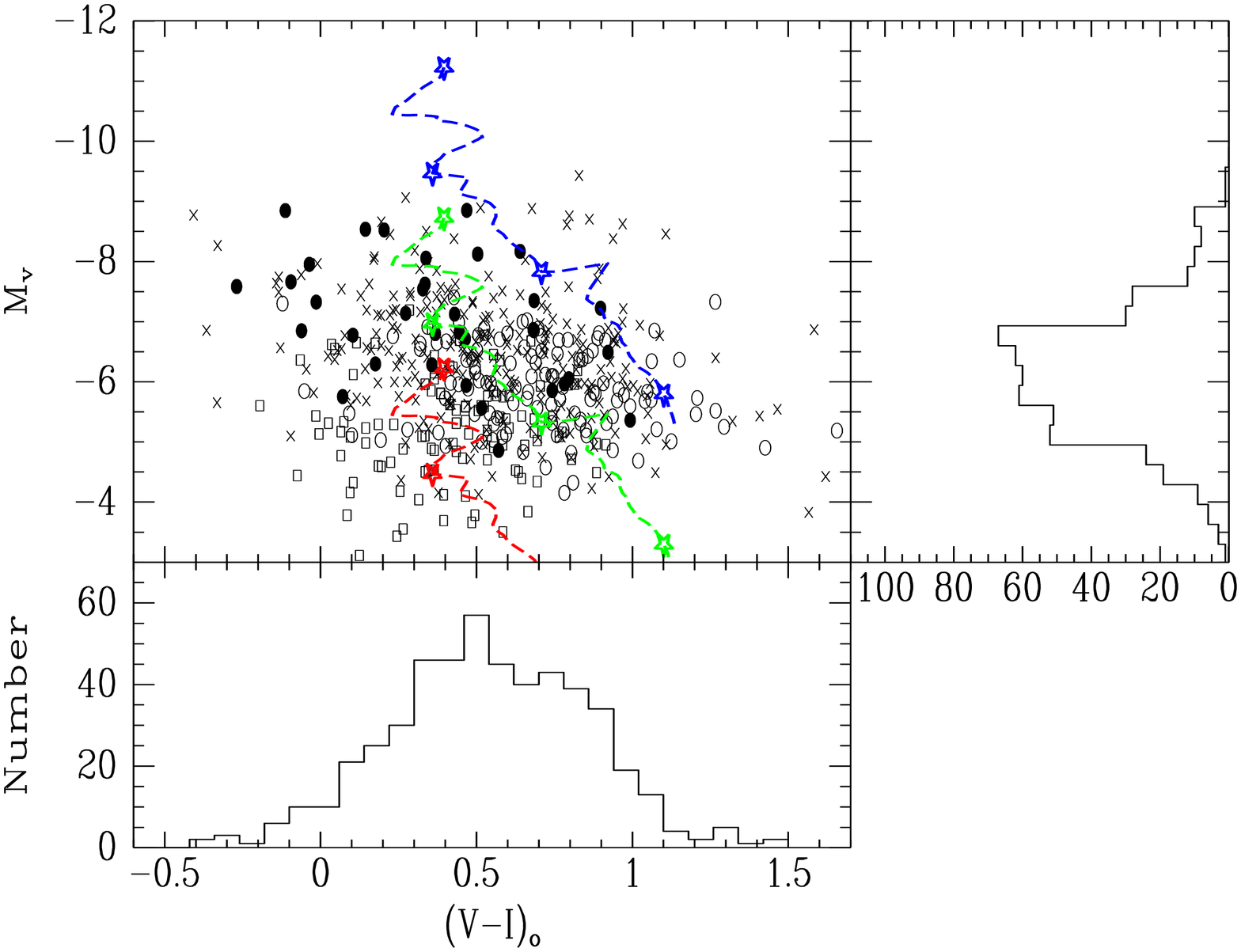}\\
\caption[]{Color-magnitude diagrams of M33 clusters. Crosses represent the clusters in this study,
filled circles represent the star clusters in
\citet{PL07}, open circles denote the star clusters in \citet{Roman09}, and open squares denote the star clusters in \citet{ZK09}. Fading lines are indicated for clusters with total initial masses of $10^5$ ({\it upper dashed line}), $10^4$, and $10^3$ ({\it lower dashed line}) $M_\odot$, assuming a Salpeter IMF (see text). Stars along
each fading line represent ages of $10^7$, $10^8$, $10^9$, and $10^{10}$ yr, from top to bottom, respectively. The vertical dashed line marks the approximate color that divides the sample star clusters into young and old populations.}
\end{figure*}

\subsection{Color-color diagram}

\begin{figure*}
\centerline{
\includegraphics[height=180mm,angle=-90]{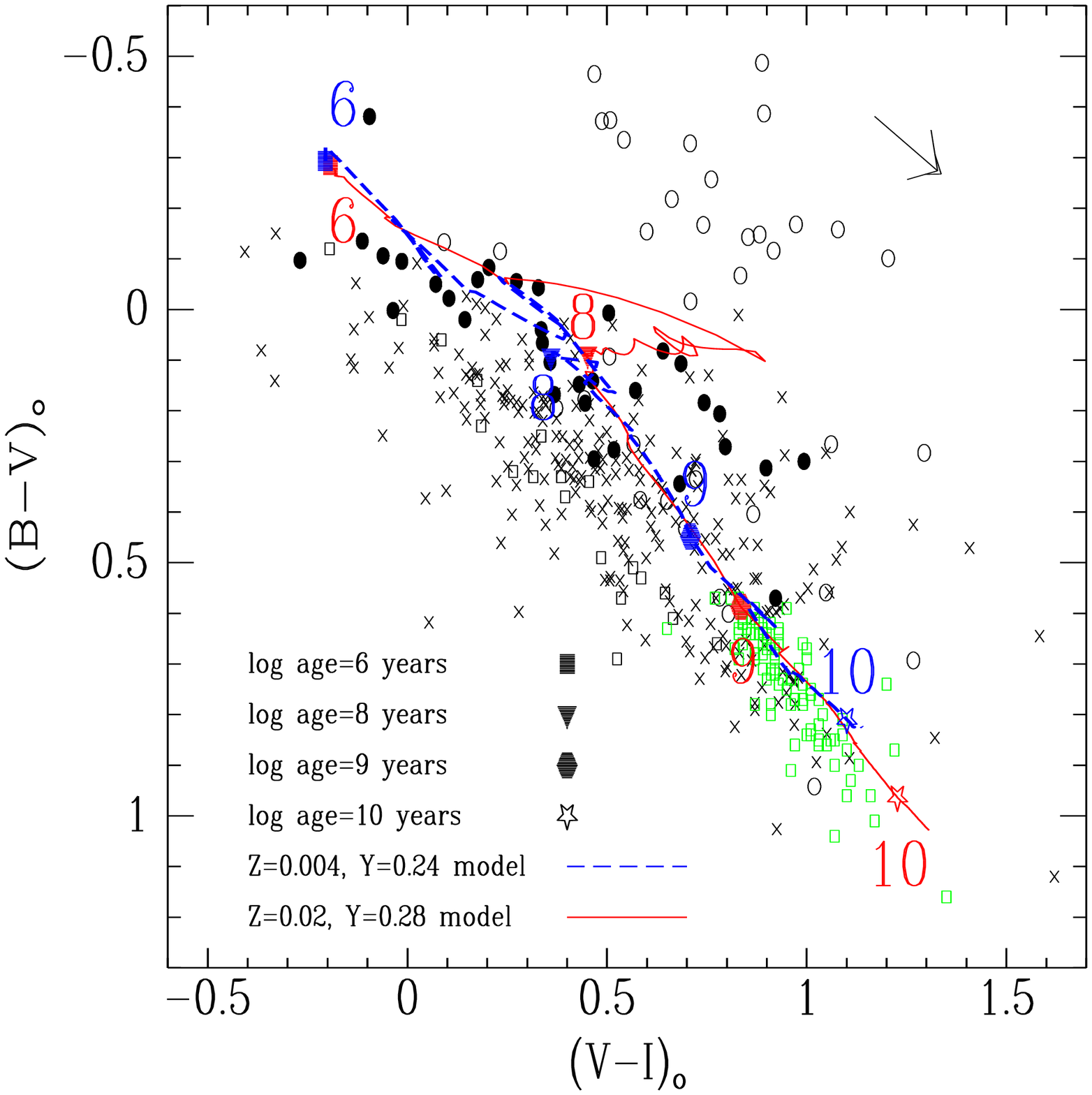}}
\caption{$(B-V)_0$ vs. $(V-I)_0$ color-color diagram of star clusters in M33. Crosses represent the clusters in this study, filled circles represent the star clusters in \citet{PL07}, open circles denote the star clusters in \citet{Roman09}, and open squares denote the star clusters in \citet{ZK09}. Green squares are Galactic globular clusters from the on-line data base of \citet{harris96} (2010 update). Theoretical evolutionary paths from the SSP model (Bruzual \& Charlot 2003) for $Z=0.004$, $Y=0.24$ (blue dashed line) and $Z=0.02$, $Y=0.28$ (red solid line) are drawn for every dex in age from $10^6$ to $10^{10}$ yr. The arrow represents the reddening direction.}
\end{figure*}

Figure 12 shows the integrated $(B-V)_0$ versus $(V-I)_0$ color-color diagram for M33 star clusters. Galactic globular clusters from the on-line data base of
Harris (1996) (2010 update) are also plotted for comparison. We overplotted the theoretical evolutionary path for the single stellar population (SSP; Bruzual \& Charlot 2003) for $Z=0.004, Y=0.24$ that was
appropriate for M33 \citep{CBF99b}. For comparison, the evolutionary path of the SSP for $Z=0.02, Y=0.28$ is also overlaid.

In general, the star clusters in M33 are located along the sequence that is consistent with the theoretical evolutionary path for $Z=0.004, Y=0.24$, while
some are on the redder or bluer side in the $(V-I)_0$ color. However, it is noted that most of sample star clusters from \citet{Roman09} are above the SSP lines. In addition, compared with the color-color diagram in \citet{PL07}, we find that the photometry in this study is shifted below the SSP lines more than in \citet{PL07}, i.e. the sample clusters in this paper are on the redder side in the $(B-V)_0$ color. In fact, from Figure 5 and Table 4, we can see that the $(B-V)$ colors obtained here are $0.127\pm 0.013$ redder than those of \citet{PL07}. From Figure 12, we also find out that, the photometry for the most Galactic globular clusters is also below the SSP lines. The continuous distribution of star clusters along the model line indicates that M33 star clusters have been formed continuously from the epoch of the first star cluster formation until recent times \citep[see also][]{PL07}. In addition, Figure 12 also shows that, some $(B-V)$ colors obtained by \citet{Roman09} are not consistent with the SSP lines.

Based on the integrated $(B-V)_0$ versus $(V-I)_0$ color-color diagram in Figure 12, we can select old globular cluster candidates in M33 which being overlapped with the Galactic globular clusters. There are $\sim 50$
star clusters which being overlapped with the Galactic globular clusters that are old globular clusters candidates in M33.

\section{Summery and conclusions}

In this paper, we present $UBVRI$ photometric measurements for 392 star clusters and cluster candidates in the field of M33 based on archival images from the LGGS \citep{massey}. These sample star clusters and cluster candidates of M33 are selected from the most recent star cluster catalog of \citet{sara07} which being compiled based on eight existing catalogs. In this catalog, the authors listed parameters such as cluster positions, magnitudes and colors in the $UBVRIJHK_s$ filters, and so on. However, a large fraction of objects in this catalog do not have previously published photometry. So, the photometric measurements in this paper supplement this catalog. Detailed comparisons show that, in general, our photometry is in agreement with previous measurements. Positions (right ascension and declination) for some clusters are corrected here.

Combined with previous literature, we obtained a large sample of M33 star clusters including 521 objects. Based on this sample of M33 star clusters, we present some statistical results:

(1) none of the M33 youngest clusters ($\sim 10^7$~yr)
have masses approaching $10^5$~$M_{\odot}$;

(2) roughly half the star clusters are consistent with the $10^4$ to $10^5$~$M_{\odot}$ mass models;

(3) the continuous distribution of star clusters along the model line indicates that M33 star clusters have been formed continuously from the epoch of the first star cluster formation until recent times;

(4) there are $\sim 50$ star clusters which being overlapped with the Galactic globular clusters on the color-color diagram, and these clusters are old globular clusters candidates in M33.

\acknowledgments
We would like to thank the anonymous referee for providing rapid and thoughtful report that helped improve the original manuscript greatly. This research was supported by the Chinese National Natural Science Foundation through grants 10873016 and 10633020, and by National Basic Research Program of China (973 Program) under grant 2007CB815403.

\clearpage
\begin{deluxetable}{ccccccc}
\tablecolumns{7} \tablewidth{0pc} \tablecaption{New $UBVRI$ photometry of M33 star clusters and candidates.\label{table1}}
\tabletypesize{\scriptsize}
\tablehead{\colhead{Object} & \colhead{$U$}  &\colhead{$B$} &\colhead{$V$}  &\colhead{$R$}   &\colhead{$I$}  &\colhead{$r_{\rm ap}$} \\
\colhead{} &\colhead{mag}    &\colhead{mag} &\colhead{mag} &\colhead{mag} &\colhead{mag} &\colhead{\arcs}
}
\startdata
1............ & 20.049 $\pm$  0.019 & 20.292 $\pm$  0.029 & 20.123 $\pm$  0.029 & 20.208 $\pm$  0.036 & 20.239 $\pm$  0.043 &  2.064\\
3............ & 21.184 $\pm$  0.053 & 21.087 $\pm$  0.064 & 20.419 $\pm$  0.042 & 20.107 $\pm$  0.041 & 19.807 $\pm$  0.037 &  2.322\\
4............ & 20.405 $\pm$  0.024 & 21.024 $\pm$  0.035 & 20.755 $\pm$  0.044 & 20.606 $\pm$  0.051 & 20.303 $\pm$  0.050 &  2.322\\
6............ & 20.223 $\pm$  0.035 & 20.276 $\pm$  0.047 & 19.767 $\pm$  0.042 & 19.444 $\pm$  0.038 & 18.970 $\pm$  0.032 &  3.096\\
7............ & 17.743 $\pm$  0.010 & 19.060 $\pm$  0.017 & 18.871 $\pm$  0.020 & 18.520 $\pm$  0.019 & 18.878 $\pm$  0.034 &  3.870\\
8............ & 20.004 $\pm$  0.028 & 19.987 $\pm$  0.034 & 19.594 $\pm$  0.029 & 19.344 $\pm$  0.025 & 19.028 $\pm$  0.023 &  2.580\\
9............ & 17.912 $\pm$  0.007 & 18.016 $\pm$  0.009 & 17.609 $\pm$  0.008 & 17.331 $\pm$  0.007 & 16.929 $\pm$  0.006 &  3.612\\
10............ & 19.741 $\pm$  0.028 & 19.597 $\pm$  0.025 & 18.780 $\pm$  0.017 & 18.294 $\pm$  0.016 & 17.676 $\pm$  0.011 &  3.096\\
\enddata
\end{deluxetable}
\addtocounter{table}{-1}
\begin{deluxetable}{ccccccc}
\tablecolumns{7} \tablewidth{0pc} \tablecaption{Continued.\label{table1}}
\tabletypesize{\scriptsize}
\tablehead{\colhead{Object} & \colhead{$U$}  &\colhead{$B$} &\colhead{$V$}  &\colhead{$R$}   &\colhead{$I$}  &\colhead{$r_{\rm ap}$} \\
\colhead{} &\colhead{mag}    &\colhead{mag} &\colhead{mag} &\colhead{mag} &\colhead{mag} &\colhead{\arcs}
}
\startdata
98............ & 18.044 $\pm$  0.009 & 18.387 $\pm$  0.012 & 18.120 $\pm$  0.012 & 17.975 $\pm$  0.015 & 17.689 $\pm$  0.016 &  2.580\\
100............ & 21.212 $\pm$  0.038 & 21.242 $\pm$  0.041 & 20.846 $\pm$  0.046 & 20.640 $\pm$  0.062 & 20.343 $\pm$  0.067 &  1.806\\
101............ & 19.138 $\pm$  0.023 & 18.832 $\pm$  0.021 & 18.010 $\pm$  0.013 & 17.518 $\pm$  0.011 & 16.917 $\pm$  0.009 &  3.612\\
102............ & 18.988 $\pm$  0.026 & 18.741 $\pm$  0.022 & 18.036 $\pm$  0.014 & 17.644 $\pm$  0.012 & 17.206 $\pm$  0.010 &  3.870\\
103............ & 19.216 $\pm$  0.016 & 19.034 $\pm$  0.015 & 18.478 $\pm$  0.014 & 18.142 $\pm$  0.017 & 17.722 $\pm$  0.015 &  3.870\\
104............ & 18.492 $\pm$  0.011 & 18.711 $\pm$  0.014 & 18.376 $\pm$  0.016 & 18.111 $\pm$  0.018 & 17.681 $\pm$  0.021 &  2.838\\
105............ & 17.983 $\pm$  0.010 & 18.380 $\pm$  0.015 & 18.040 $\pm$  0.016 & 17.752 $\pm$  0.016 & 17.125 $\pm$  0.014 &  2.580\\
106............ & 19.034 $\pm$  0.017 & 18.988 $\pm$  0.020 & 18.650 $\pm$  0.021 & 18.387 $\pm$  0.024 & 18.135 $\pm$  0.027 &  2.580\\
\enddata
\end{deluxetable}
\addtocounter{table}{-1}
\begin{deluxetable}{ccccccc}
\tablecolumns{7} \tablewidth{0pc} \tablecaption{Continued.\label{table1}}
\tabletypesize{\scriptsize}
\tablehead{\colhead{Object} & \colhead{$U$}  &\colhead{$B$} &\colhead{$V$}  &\colhead{$R$}   &\colhead{$I$}  &\colhead{$r_{\rm ap}$} \\
\colhead{} &\colhead{mag}    &\colhead{mag} &\colhead{mag} &\colhead{mag} &\colhead{mag} &\colhead{\arcs}
}
\startdata
185............ & 20.203 $\pm$  0.019 & 20.930 $\pm$  0.036 & 20.704 $\pm$  0.044 & 20.583 $\pm$  0.053 & 20.406 $\pm$  0.079 &  1.806\\
186............ & 19.478 $\pm$  0.020 & 19.336 $\pm$  0.020 & 18.625 $\pm$  0.018 & 18.236 $\pm$  0.018 & 17.741 $\pm$  0.015 &  2.580\\
187............ & 20.636 $\pm$  0.028 & 20.687 $\pm$  0.031 & 20.261 $\pm$  0.032 & 19.966 $\pm$  0.039 & 19.332 $\pm$  0.035 &  1.548\\
188............ & 18.677 $\pm$  0.018 & 18.552 $\pm$  0.017 & 17.665 $\pm$  0.011 & 17.135 $\pm$  0.008 & 16.643 $\pm$  0.007 &  3.870\\
189............ & 20.587 $\pm$  0.039 & 20.582 $\pm$  0.040 & 19.922 $\pm$  0.035 & 19.511 $\pm$  0.036 & 19.024 $\pm$  0.036 &  2.322\\
190............ & 18.197 $\pm$  0.008 & 18.424 $\pm$  0.011 & 18.132 $\pm$  0.011 & 17.965 $\pm$  0.013 & 17.650 $\pm$  0.016 &  2.580\\
191............ & 18.917 $\pm$  0.014 & 18.800 $\pm$  0.014 & 18.316 $\pm$  0.015 & 18.037 $\pm$  0.015 & 17.654 $\pm$  0.015 &  2.322\\
192............ & 19.841 $\pm$  0.044 & 19.831 $\pm$  0.036 & 19.131 $\pm$  0.032 & 18.765 $\pm$  0.034 & 18.139 $\pm$  0.033 &  2.838\\
\enddata
\end{deluxetable}
\addtocounter{table}{-1}
\begin{deluxetable}{ccccccc}
\tablecolumns{7} \tablewidth{0pc} \tablecaption{Continued.\label{table1}}
\tabletypesize{\scriptsize}
\tablehead{\colhead{Object} & \colhead{$U$}  &\colhead{$B$} &\colhead{$V$}  &\colhead{$R$}   &\colhead{$I$}  &\colhead{$r_{\rm ap}$} \\
\colhead{} &\colhead{mag}    &\colhead{mag} &\colhead{mag} &\colhead{mag} &\colhead{mag} &\colhead{\arcs}
}
\startdata
269............ & 17.807 $\pm$  0.007 & 18.087 $\pm$  0.010 & 17.756 $\pm$  0.011 & 17.595 $\pm$  0.016 & 17.353 $\pm$  0.021 &  3.096\\
270............ & 18.347 $\pm$  0.014 & 18.412 $\pm$  0.014 & 17.930 $\pm$  0.013 & 17.635 $\pm$  0.014 & 17.276 $\pm$  0.014 &  3.354\\
271............ & 18.976 $\pm$  0.018 & 19.031 $\pm$  0.021 & 18.528 $\pm$  0.020 & 18.159 $\pm$  0.020 & 17.686 $\pm$  0.020 &  3.354\\
272............ & 18.574 $\pm$  0.025 & 18.819 $\pm$  0.027 & 18.512 $\pm$  0.030 & 18.417 $\pm$  0.041 & 18.083 $\pm$  0.053 &  3.354\\
273............ & 17.346 $\pm$  0.007 & 17.198 $\pm$  0.006 & 16.399 $\pm$  0.005 & 15.922 $\pm$  0.005 & 15.414 $\pm$  0.006 &  3.354\\
274............ & 18.202 $\pm$  0.009 & 19.119 $\pm$  0.018 & 18.956 $\pm$  0.020 & 18.525 $\pm$  0.019 & 18.377 $\pm$  0.024 &  2.838\\
275............ & 17.815 $\pm$  0.008 & 17.521 $\pm$  0.008 & 16.545 $\pm$  0.006 & 15.930 $\pm$  0.005 & 15.313 $\pm$  0.005 &  3.354\\
276............ & 17.566 $\pm$  0.019 & 18.355 $\pm$  0.016 & 18.186 $\pm$  0.017 & 18.110 $\pm$  0.030 & 18.125 $\pm$  0.025 &  1.548\\
\enddata
\end{deluxetable}
\addtocounter{table}{-1}
\begin{deluxetable}{ccccccc}
\tablecolumns{7} \tablewidth{0pc} \tablecaption{Continued.\label{table1}}
\tabletypesize{\scriptsize}
\tablehead{\colhead{Object} & \colhead{$U$}  &\colhead{$B$} &\colhead{$V$}  &\colhead{$R$}   &\colhead{$I$}  &\colhead{$r_{\rm ap}$} \\
\colhead{} &\colhead{mag}    &\colhead{mag} &\colhead{mag} &\colhead{mag} &\colhead{mag} &\colhead{\arcs}
}
\startdata
361............ & 20.099 $\pm$  0.024 & 20.235 $\pm$  0.033 & 19.857 $\pm$  0.036 & 19.474 $\pm$  0.038 & 18.787 $\pm$  0.032 &  2.580\\
364............ & 20.018 $\pm$  0.036 & 19.688 $\pm$  0.028 & 19.195 $\pm$  0.027 & 18.808 $\pm$  0.026 & 18.383 $\pm$  0.026 &  3.612\\
365............ & 19.285 $\pm$  0.020 & 19.486 $\pm$  0.023 & 19.267 $\pm$  0.027 & 19.147 $\pm$  0.035 & 19.080 $\pm$  0.046 &  2.580\\
366............ & 19.022 $\pm$  0.013 & 20.023 $\pm$  0.026 & 20.022 $\pm$  0.035 & 19.758 $\pm$  0.036 & 19.926 $\pm$  0.055 &  2.580\\
367............ & 21.749 $\pm$  0.119 & 21.044 $\pm$  0.066 & 19.459 $\pm$  0.021 & 18.620 $\pm$  0.015 & 17.869 $\pm$  0.012 &  1.548\\
368............ & 16.646 $\pm$  0.003 & 17.000 $\pm$  0.005 & 16.685 $\pm$  0.005 & 16.481 $\pm$  0.005 & 16.201 $\pm$  0.005 &  3.096\\
370............ & 16.690 $\pm$  0.004 & 17.042 $\pm$  0.005 & 16.738 $\pm$  0.005 & 16.559 $\pm$  0.005 & 16.269 $\pm$  0.005 &  3.096\\
371............ & 17.037 $\pm$  0.004 & 17.387 $\pm$  0.006 & 17.123 $\pm$  0.006 & 16.978 $\pm$  0.007 & 16.765 $\pm$  0.007 &  3.096\\
\enddata
\end{deluxetable}

\clearpage
\begin{deluxetable}{cccccccccccc}
\tablecolumns{8} \tablewidth{0pc} \tablecaption{Comparison of $V$ photometry for M33 star clusters and cluster candidates considered here between this study
and previous studies.\label{table2}}
\tabletypesize{\scriptsize}
\tablehead{\colhead{ID\tablenotemark{a}} & \colhead{ID\tablenotemark{b}} & \colhead{ID\tablenotemark{c}}  &\colhead{$V$\tablenotemark{d}} &\colhead{$V$\tablenotemark{e}}  &\colhead{$V$\tablenotemark{f}} &\colhead{$V$\tablenotemark{g}}  &\colhead{$\Delta V$\tablenotemark{h}}  & \colhead{$\Delta V$\tablenotemark{i}}  & \colhead{$\Delta V$\tablenotemark{j}} & \colhead{$r_{\rm ap}$\tablenotemark{k}} \\
&              &               &\colhead{(mag)}&\colhead{(mag)}&\colhead{(mag)}&\colhead{(mag)}&\colhead{(mag)} &\colhead{(mag)}&  \colhead{(mag)}&\colhead{\arcs}}
\startdata
   44......&.........&.........&  19.04 &  ..............& ..... &20.180$\pm$0.038 &   1.14 &   .....&.....&      1.290\\
   45......&111......&.........&  19.56 &18.884$\pm$0.013& ..... &20.170$\pm$0.032 &   0.61 &  1.286 &.....&      1.290\\
   46......&.........&.........&  18.27 &  ..............& ..... &22.556$\pm$0.163 &   4.29 &   .....&.....&      1.032\\
   66......&34...... &36.......&  18.65 &18.090$\pm$0.010& 18.339&19.249$\pm$0.026 &   0.60 &  1.159 &0.910&      1.548\\
  116......&.........&.........&  18.52 &  ..............& ..... &21.174$\pm$0.116 &   2.65 &   .....&.....&      1.548\\
  118......&.........&66.......&  18.84 &  ..............& 18.835&20.077$\pm$0.039 &   1.24 &   .....&1.242&      1.290\\
  120......&.........&67.......&  19.19 &  ..............& 19.395&20.262$\pm$0.050 &   1.07 &   .....&0.867&      1.806\\
  153......&139......&.........&  18.57 &18.088$\pm$0.019& ..... &20.047$\pm$0.051 &   1.48 &  1.959 &.....&      1.290\\
\enddata
\tablenotetext{a}{The object names following the naming convention of \citet{sara07}.}
\tablenotetext{b}{The object names following the naming convention of \citet{PL07}.}
\tablenotetext{c}{The object names following the naming convention of \citet{Roman09}.}
\tablenotetext{d}{The photometric data being collected in \citet{sara07}.}
\tablenotetext{e}{The photometry obtained in \citet{PL07}.}
\tablenotetext{f}{The photometry obtained in \citet{Roman09}.}
\tablenotetext{g}{The photometry obtained in this paper.}
\tablenotetext{h}{The magnitude difference between this study and \citet{sara07} (this study minus
Sarajedini \& Mancone 2007).}
\tablenotetext{i}{The magnitude difference between this study and \citet{PL07} (this study minus
Park \& Lee 2007).}
\tablenotetext{j}{The magnitude difference between this study and \citet{Roman09} (this study minus
San Roman et al. 2009).}
\tablenotetext{k}{The aperture radius of photometry adopted in this paper.}
\end{deluxetable}

\begin{deluxetable}{lccc}
\tablecolumns{4} \tablewidth{0pc} \tablecaption{Comparisons of our broadband photometry with previously published M33 star cluster photometry being collected in \citet{sara07}. Offset = our measurement $-$ literature value.\label{table3}}
\tabletypesize{\scriptsize}
\tablehead{\colhead{Band} &\colhead{ Mean Offset } &\colhead{$\sigma$}
&\colhead{Number} }
\startdata
  $V$   & $  0.187\pm 0.028$ &  0.462 & 265 \\
  $B-V$ & $  0.041\pm 0.017$ &  0.221 & 173 \\
  $V-I$ & $  0.004\pm 0.020$ &  0.257 & 171 \\
\enddata
\end{deluxetable}

\begin{deluxetable}{lccc}
\tablecolumns{4} \tablewidth{0pc} \tablecaption{Comparisons of our broadband photometry with previously published M33 star cluster photometry in \citet{PL07}. Offset = our measurement $-$ literature value.\label{table4}}
\tabletypesize{\scriptsize}
\tablehead{\colhead{Band} &\colhead{ Mean Offset } &\colhead{$\sigma$}
&\colhead{Number} }
\startdata
  $V$   & $  0.195\pm 0.029$ &  0.398 & 191 \\
  $B-V$ & $  0.127\pm 0.012$ &  0.167 & 180 \\
  $V-I$ & $ -0.011\pm 0.014$ &  0.195 & 189 \\
\enddata
\end{deluxetable}

\begin{deluxetable}{lccc}
\tablecolumns{5} \tablewidth{0pc} \tablecaption{Comparisons of our broadband photometry with previously published M33 star cluster photometry in \citet{Roman09}. Offset = our measurement $-$ literature value.\label{table5}}
\tabletypesize{\scriptsize}
\tablehead{\colhead{Band} &\colhead{ Mean Offset } &\colhead{$\sigma$}
&\colhead{Number} }
\startdata
  $V$   & $  0.220\pm 0.049$ &  0.318 & 43 \\
  $B-V$ & $  0.416\pm 0.068$ &  0.265 & 16 \\
  $V-I$ & $ -0.060\pm 0.045$ &  0.294 & 43 \\
\enddata
\end{deluxetable}

\begin{deluxetable}{lccc}
\tablecolumns{4} \tablewidth{0pc} \tablecaption{Comparisons of our broadband photometry with previously published M33 star cluster photometry being collected in \citet{sara07}. Offset = our measurement $-$ literature value. Photometries of M33 star clusters considered here are derived with an aperture of $r=3.354''$ (13 pixels).\label{table6}}
\tabletypesize{\scriptsize}
\tablehead{\colhead{Band} &\colhead{ Mean Offset } &\colhead{$\sigma$}
&\colhead{Number} }
\startdata
  $V$   & $  0.026\pm 0.023$ &  0.372 & 265 \\
  $B-V$ & $  0.022\pm 0.016$ &  0.210 & 173 \\
  $V-I$ & $  0.084\pm 0.024$ &  0.313 & 171 \\
\enddata
\end{deluxetable}

\begin{deluxetable}{lccc}
\tablecolumns{4} \tablewidth{0pc} \tablecaption{Comparisons of our broadband photometry with previously published M33 star cluster photometry in \citet{PL07}. Offset = our measurement $-$ literature value. Photometries of M33 star clusters considered here are derived with an aperture of $r=3.354''$ (13 pixels).\label{table7}}
\tabletypesize{\scriptsize}
\tablehead{\colhead{Band} &\colhead{ Mean Offset } &\colhead{$\sigma$}
&\colhead{Number} }
\startdata
  $V$   & $  0.021\pm 0.019$ &  0.265 & 191 \\
  $B-V$ & $  0.107\pm 0.011$ &  0.154 & 180 \\
  $V-I$ & $  0.057\pm 0.017$ &  0.230 & 189 \\
\enddata
\end{deluxetable}

\begin{deluxetable}{lccc}
\tablecolumns{5} \tablewidth{0pc} \tablecaption{Comparisons of our broadband photometry with previously published M33 star cluster photometry in \citet{Roman09}. Offset = our measurement $-$ literature value. Photometries of M33 star clusters considered here are derived with an aperture of $r=3.354''$ (13 pixels).\label{table8}}
\tabletypesize{\scriptsize}
\tablehead{\colhead{Band} &\colhead{ Mean Offset } &\colhead{$\sigma$}
&\colhead{Number}}
\startdata
  $V$   & $ -0.008\pm 0.046$ &  0.295 & 43 \\
  $B-V$ & $  0.405\pm 0.068$ &  0.263 & 16 \\
  $V-I$ & $ -0.023\pm 0.038$ &  0.249 & 43 \\
\enddata
\end{deluxetable}

\begin{deluxetable}{cccccccccccc}
\tablecolumns{8} \tablewidth{0pc} \tablecaption{Comparison of photometry in $V$ band between this study
and previous studies. Photometries of M33 star clusters considered here are derived with an aperture of $r=3.354''$ (13 pixels).\label{table9}}
\tabletypesize{\scriptsize}
\tablehead{\colhead{ID\tablenotemark{a}} & \colhead{ID\tablenotemark{b}} & \colhead{ID\tablenotemark{c}}  &\colhead{$V$\tablenotemark{d}} &\colhead{$V$\tablenotemark{e}}  &\colhead{$V$\tablenotemark{f}} &\colhead{$V$\tablenotemark{g}}  &\colhead{$\Delta V$\tablenotemark{h}}  & \colhead{$\Delta V$\tablenotemark{i}}  & \colhead{$\Delta V$\tablenotemark{j}} & \colhead{$r_{\rm ap}$\tablenotemark{k}} \\
&              &               &\colhead{(mag)}&\colhead{(mag)}&\colhead{(mag)}&\colhead{(mag)}&\colhead{(mag)} &\colhead{(mag)}&  \colhead{(mag)}&\colhead{\arcs}}
\startdata
   46......&.........&.........&  18.27 &  ..............& ..... &20.934$\pm$0.158 &   2.66 &   .....&.....&      3.354\\
   90......&117......&.........&  19.65 &19.629$\pm$0.007& ......&17.704$\pm$0.007 &$-1.95$ &$-1.925$&.....&      3.354\\
  132......&133......&74.......&  17.71 &16.602$\pm$0.004& 17.190&16.559$\pm$0.007 &$-1.15$ &$-0.043$&$-0.631$&   3.354\\
  193......&157......&.........&  18.49 &19.878$\pm$0.053& ..... &20.088$\pm$0.082 &   1.60 &  0.210 &.....&      3.354\\
  195......&158......&.........&  18.84 &18.246$\pm$0.023& ..... &19.638$\pm$0.126 &   0.80 &  1.392 &.....&      3.354\\
  196......&159......&.........&  19.27 &  ..............& ..... &20.857$\pm$0.125 &   1.60 &   .....&.....&      3.354\\
  231......&69...... &137......&  18.26 &17.410$\pm$0.006& 18.702&17.504$\pm$0.010 &$-0.76$ &  0.094 &$-1.198$&   3.354\\
  236......&72.......&.........&  18.75 &20.146$\pm$0.117& ..... &18.939$\pm$0.051 &   0.19 &$-1.207$&.....&      3.354\\
\enddata
\tablenotetext{a}{The object names following the naming convention of \citet{sara07}.}
\tablenotetext{b}{The object names following the naming convention of \citet{PL07}.}
\tablenotetext{c}{The object names following the naming convention of \citet{Roman09}.}
\tablenotetext{d}{The photometric data being collected in \citet{sara07}.}
\tablenotetext{e}{The photometry obtained in  \citet{PL07}.}
\tablenotetext{f}{The photometry obtained in \citet{Roman09}.}
\tablenotetext{g}{The photometry obtained in this paper.}
\tablenotetext{h}{The magnitude difference between this study and \citet{sara07} (this study minus
Sarajedini \& Mancone 2007).}
\tablenotetext{i}{The magnitude difference between this study and \citet{PL07} (this study minus
Park \& Lee 2007).}
\tablenotetext{j}{The magnitude difference between this study and \citet{Roman09} (this study minus
San Roman et al. 2009).}
\tablenotetext{k}{The aperture radius of photometry adopted in this paper.}
\end{deluxetable}

\end{document}